\documentclass[conference]{IEEEtran}
\ifCLASSINFOpdf
\else
\fi

\usepackage{etoolbox}
\apptocmd{\thebibliography}{\setlength{\itemsep}{4pt}}{}{}

\PassOptionsToPackage{hyphens}{url}\usepackage[hidelinks]{hyperref}

\usepackage{enumitem}
\usepackage{graphicx}
\usepackage[flushleft]{threeparttable}
\usepackage{array}

\usepackage{multirow}

\newcommand{\dataset}{\textsc{DroidLeaks}\ }
\newcommand{\datasetns}{\textsc{DroidLeaks}}
\newcommand{\numbugs}{298\ }
\newcommand{\numbugsns}{298}

\begin{document}
%
\title{\textsc{DroidLeaks}: Benchmarking Resource Leak Bugs for Android Applications}

\author{\IEEEauthorblockN{Yepang Liu\IEEEauthorrefmark{2},
		Lili Wei\IEEEauthorrefmark{2}, Chang Xu\IEEEauthorrefmark{4} and
		Shing-Chi Cheung\IEEEauthorrefmark{2}}
	\IEEEauthorblockA{\IEEEauthorrefmark{2}Dept. of Computer Science and Engineering, The Hong Kong Univ. of Science and Technology, Hong Kong, China\\
		\IEEEauthorrefmark{4}State Key Lab for Novel Software Tech. and Dept. of Comp. Sci. and Tech., Nanjing University, Nanjing, China\\
		\IEEEauthorrefmark{2}\{andrewust, lweiae, scc\}@cse.ust.hk,
		\IEEEauthorrefmark{4}changxu@nju.edu.cn}}

\maketitle

\begin{abstract}
Resource leak bugs in Android apps are pervasive and can cause serious performance degradation and system crashes. In recent years, several resource leak detection techniques have been proposed to assist Android developers in correctly managing system resources. Yet, there exist no common bug benchmarks for effectively and reliably comparing such techniques and quantitatively evaluating their strengths and weaknesses. This paper describes our initial contribution towards constructing such a benchmark. To locate real resource leak bugs, we mined 124,215 code revisions of 34 large-scale open-source Android apps. We successfully found \numbugs fixed resource leaks, which cover a diverse set of resource classes, from 32 out of the 34 apps. To understand the characteristics of these bugs, we conducted an empirical study, which revealed the root causes of frequent resource leaks in Android apps and common patterns of faults made by developers. With our findings, we further implemented a static checker to detect a common pattern of resource leaks in Android apps. Experiments showed that the checker can effectively locate real resource leaks in popular Android apps, confirming the usefulness of our work.

\vspace{6pt}
\end{abstract}

\IEEEpeerreviewmaketitle

\section{Introduction}
Mobile applications (apps) such as those running on Android platforms are gaining their popularity in recent years. The user downloads of such apps have long surpassed that of PC software. However, unlike PC software, mobile apps run on resource-constrained devices and are expected to consume computational resources (e.g., memory, battery power) more efficiently. However, many apps on the market often do not properly release their acquired computational resources after use~\cite{relda}. Such defects are called \textit{resource leaks} and they gradually deplete the finite computational resources at runtime, leading to severe performance degradation and system crashes. 

Ensuring proper resource usage in a program is a non-trivial task for developers~\cite{tracker}. Over the years, both research communities and industries have spent much effort in developing automated techniques to help mobile developers manage resources used by their apps. For example, Wu et al. proposed a light-weight static analysis technique to detect potential leaks of resources in Android apps~\cite{relda}\cite{relda2}. Vekris et al. and our earlier work proposed data flow analysis~\cite{verifier}\cite{elite} and model checking techniques~\cite{greendroid}\cite{greendroid_tse} to detect the leak of wake locks, a special type of system resource for keeping device awake to perform long running operations, in Android apps. Besides, tools developed by industries such as Facebook Infer~\cite{infer} and the built-in checkers in Eclipse~\cite{jdt} and Android Studio~\cite{studio} can also help developers find resource leak bugs in their Android or iOS apps.

Despite the tremendous efforts made towards automated resource management and leak detection, there does not exist a widely-recognized benchmark of resource leak bugs for mobile apps. Such benchmarks are essential as they provide a common and reliable basis based on which one can evaluate and compare various resource management and leak detection techniques. Due to the lack of such benchmarks, the authors of existing techniques~\cite{relda}\cite{greendroid_tse}\cite{elite}\cite{verifier}\cite{relda2} had to conduct experiments on their own selected subjects (usually a small set). As a result, it is hard to quantitatively compare such techniques' strengths and limitations to understand the state of the art and push the area forward.

In this work, we aim to make an initial contribution towards benchmarking resource leak bugs for mobile apps. We focus on the Android platform. To collect real resource leak bugs in Android apps, we investigated 34 diverse, large-scale, well-maintained, and popular open-source Android apps indexed by F-Droid~\cite{fdroid}. A straightforward approach for bug collection would be searching these apps' issue tracking system. However, in practice, this approach can miss many resource leak bugs that are fixed without being documented. Fortunately, patches to fix bugs found during the development process or reported by users are eventually committed to the apps' code repository. Therefore, we can resort to code repository mining to collect bugs. Specifically, by automatically mining 124,215 code revisions of our app subjects and careful manual checking, we successfully located \numbugs fixed resource leak bugs in 32 out of the 34 apps, of which only 14 (4.7\% = 14 / \numbugsns) are documented in the issue tracking systems. We call this collection of bugs \dataset and collected the following data for each bug: (1) the name of the infected app, (2) the type of the leaked system resource, (3) bug location and the buggy code revision, (4) bug-fixing patches, and (5) bug report (if located). To understand the characteristics of these bugs, we further performed an empirical study on \dataset and studied the following three research questions:

\begin{itemize}[leftmargin=*, itemsep=4pt]
	\item{\textbf{RQ1 (Resource type and consequence of leak):} \textit{What types of system resources are leaked due to these bugs? What are the consequences of these resource leaks? Are the leaked resources specific to the Android platform?}}
	
	\item{\textbf{RQ2 (Resource leak extent):} \textit{Did the developers completely forget to release the concerned resources or only forget to release the resources on certain program execution or exceptional paths?}}
	
	\item{\textbf{RQ3 (Common fault patterns):} \textit{Are there common patterns of faults made by developers, which lead to resource leaks?}}
\end{itemize}

The study led to several interesting findings. For example, we found that due to the complex control flows caused by the event-driven computing paradigm, Android developers completely forgot to release acquired system resources for 191 of the \numbugs bugs (64.1\%). In comparison, resource leaks on exceptional paths are minority (56 out of \numbugsns) in \datasetns. We also identified five root causes (e.g., unexpected user interactions and environment interplay) for the prevalence of resource leaks in Android apps and found three common patterns of faults made by developers (e.g., API misuses and losing references to resource objects). Moreover, we observed that bugs in \dataset are representative and comprehensive as they cover all types of resource leaks that were used to evaluate the latest resource leak detection techniques for Android apps~\cite{relda}\cite{relda2}\cite{verifier}\cite{elite}, and additionally contain more types. Such findings suggest that our work not only can provide practical programming guidance to Android developers but also can support follow-up research on developing automated bug finding and patching techniques.

As a large collection of real bugs, \dataset has many potential applications. In this work, we further demonstrated an application of \datasetns. We conducted a case study to investigate whether our findings obtained by studying bugs in \dataset can help resource leak detection. Specifically, we implemented a light-weight static code analyzer, which checks for the existence of our observed resource leak bug pattern, and applied it on the latest version of our 34 app subjects. Encouragingly, the checker successfully found 18 resource leak bugs in nine apps, most of which were later quickly confirmed and fixed by developers. In summary, our work makes three major contributions:

\begin{itemize}[leftmargin=*, itemsep=4pt]
	\item{We present \datasetns, a collection of real resource leak bugs found in representative open-source Android apps. Its initial version described in this paper features \numbugs bugs covering 37 different resource classes. This collection of bugs can facilitate the future development of robust techniques for resource leak finding and patching. To the best of our knowledge, \dataset is the first of its kind and we will release it for public access~\cite{leakbench}.}
	\item{We performed an empirical study on \datasetns. The study revealed the root causes why resource leak bugs are prevalent in Android apps and found common patterns of coding mistakes made by developers.}
	\item{We conducted a real case study of in-the-wild bug detection by leveraging our empirical findings. The study found real resource leak bugs in popular Android apps, confirming the usefulness of \datasetns.}
\end{itemize}

The rest of this paper is organized as follows. Section~\ref{sec:bg} introduces the preliminaries of Android apps and resource leak bugs. Section~\ref{sec:method} presents our approach to constructing \datasetns. Section~\ref{sec:empirical} discusses the characteristics of bugs documented in \datasetns. Section~\ref{sec:case_study} reports our bug detection case study. Section~\ref{sec:related_work} discusses related work and Section~\ref{sec:conclusion} concludes this paper.

\section{Background}\label{sec:bg}

Android is a Linux-based open-source mobile operating system. The apps running on Android platforms are mostly written in Java and compiled to Dalvik bytecode, which are then encapsulated into Android app package files (i.e., \texttt{\small .apk} files) for distribution and installation~\cite{api}.

\textbf{App components and event handlers}. Android apps are event-driven programs. An app usually consists of four types of components: (1) \textit{activities} contain graphical user interfaces for user interactions, (2) \textit{services} run at background for long-running operations, (3) \textit{broadcast receivers} respond to system-wide broadcast messages, and (4) \textit{content providers} manage shared app data for queries. Each app component can define and register a set of \textit{event handlers}: callbacks that will be invoked by the Android OS when certain events occur. These event handlers implement the main logic of an app.

\textbf{Resource management}. To request system resources (e.g., camera) for computation, Android apps can invoke designated APIs (e.g., \texttt{\small android.hardware.Camera.open()}) defined in the standard Android SDK, JDK, or third-party libraries. When the computation completes, the apps should release the acquired resources by invoking the corresponding pairing APIs (e.g., \texttt{\small android.hardware.Camera.release()}). For correct resource management, developers need to make sure that such resource releasing is performed on every possible program execution path, including exceptional ones. Particularly, for reference counted resources (e.g., in Android, wake locks are by default reference counted~\cite{api}), developers need to balance the number of calls to the resource releasing API and that to the resource acquiring API. Otherwise, the resources will be leaked, which can cause undesirable consequences such as performance degradation and system crashes. In practice, such resource management tasks are error-prone. The complex control flows among Android event handlers further complicate the tasks, giving rise to various resource leak bugs.

\section{Methodology}\label{sec:method}

\begin{table*}[htbp!]
	\begin{threeparttable}
		\centering
		\caption{Open-Source App Subjects and Their Resource Leak Bugs}
		\renewcommand{\arraystretch}{1.15}
		\label{tab:subjects}
		\begin{tabular}{|c|c|c|c|c|c|c|c|}
			\hline
			\textbf{App name} & \textbf{Category} & \textbf{Rating} & \textbf{Downloads} & \textbf{SLOC (Java)} & \textbf{\# total revisions} & \textbf{\# interesting revisions} & \textbf{\# bugs} \\ \hline
			AnkiDroid & Education & 4.5 & 1M -- 5M & 47.3K & 8,303 & 223 & 30 \\ \hline
			AnySoftKeyboard & Tools & 4.4 & 1M -- 5M & 25.2K & 2,803 & 46 & 4 \\ \hline
			APG & Communication & 4.4 & 100K -- 500K & 42.0K & 4,366 & 69 & 10 \\ \hline
			BankDroid & Finance & 4.1 & 100K -- 500K & 22.9K & 1,202 & 5 & 6 \\ \hline
			Barcode Scanner & Shopping & 4.1 & 100M -- 500M & 10.6K & 3,219 & 43 & 3 \\ \hline
			BitCoin Wallet & Finance & 4.0 & 1M -- 5M & 18.0K & 2,442 & 52 & 4 \\ \hline
			CallMeter & Tools & 4.3 & 1M -- 5M & 13.5K & 2,263 & 27 & 10 \\ \hline
			ChatSecure & Communication & 4.0 & 500K -- 1M & 37.2K & 2,906 & 128 & 32 \\ \hline
			ConnectBot & Communication & 4.6 & 1M -- 5M & 17.6K & 1,349 & 22 & 5 \\ \hline
			CSipSimple & Communication & 4.3 & 1M -- 5M & 49.0K & 1,778 & 42 & 7 \\ \hline
			CycleStreets & Travel \& Local & 3.7 & 50K -- 100K & 18.8K & 1,269 & 18 & 1 \\ \hline
			c:geo & Entertainment & 4.4 & 1M -- 5M & 52.2K & 9,338 & 90 & 8 \\ \hline
			FBReader & Books \& References & 4.5 & 10M -- 50M & 70.9K & 9,005 & 76 & 8 \\ \hline
			Google Authenticator & Tools & 4.4 & 10M -- 50M & 3.3K & 179 & 1 & 5 \\ \hline
			Hacker News Reader & News \& Magazines & 4.4 & 50K -- 100K & 4.0K & 296 & 2 & 4 \\ \hline
			IRCCloud & Communication & 4.3 & 50K -- 100K & 35.3K & 1,866 & 136 & 14 \\ \hline
			K-9 Mail & Communication & 4.3 & 5M -- 10M & 78.5K & 6,132 & 98 & 31 \\ \hline
			OI File Manager & Productivity & 4.2 & 5M -- 10M & 6.9K & 399 & 9 & 0 \\ \hline
			Open GPS Tracker & Travel \& Local & 4.1 & 500K -- 1M & 12.3K & 1,096 & 42 & 2 \\ \hline
			Osmand & Maps \& Navigation & 4.2 & 1M -- 5M & 137.7K & 29,336 & 134 & 13 \\ \hline
			OsmDroid & Maps \& Navigation & 3.9 & 50K -- 100K & 18.4K & 1,881 & 34 & 2 \\ \hline
			OSMTracker & Travel \& Local & 4.4 & 100K -- 500K & 5.9K & 400 & 14 & 4 \\ \hline
			ownCloud & Productivity & 3.7 & 100K -- 500K & 31.6K & 4,541 & 82 & 8 \\ \hline
			Quran for Android & Books \& References & 4.7 & 10M -- 50M & 21.7K & 1,560 & 47 & 19 \\ \hline
			SipDroid & Communication & 4.0 & 1M -- 5M & 24.5K & 293 & 7 & 1 \\ \hline
			SMSDroid & Communication & 3.9 & 500K -- 1M & 4.7K & 813 & 14 & 2 \\ \hline
			SureSpot & Social & 4.2 & 100K -- 500K & 41.0K & 1,572 & 77 & 12 \\ \hline
			Terminal Emulator & Tools & 4.4 & 10M -- 50M & 11.7K & 1,035 & 11 & 2 \\ \hline
			Transdroid & Tools & 4.4 & 100K -- 500K & 23.5K & 427 & 8 & 2 \\ \hline
			Ushahidi & Communication & 3.7 & 10K -- 50K & 35.7K & 948 & 16 & 4 \\ \hline
			VLC & Video Players \& Editors & 4.3 & 10M -- 50M & 18.1K & 3,481 & 60 & 14 \\ \hline
			WebSMS & Communication & 4.4 & 100K -- 500K & 4.4K & 1,648 & 10 & 0 \\ \hline
			WordPress & Social & 4.2 & 5M -- 10M & 74.9K & 14,805 & 162 & 29 \\ \hline
			Xabber & Communication & 4.2 & 500K -- 1M & 38.2K & 1,264 & 6 & 2 \\ \hline
		\end{tabular}
		\begin{tablenotes}
			\item Notes: (1) 1K = 1,000 and 1M = 1,000,000; (2) For app downloads, we only considered data from the Google Play store.
		\end{tablenotes}
	\end{threeparttable}
\end{table*}

\subsection{Selecting Open-Source App Subjects}\label{ssec:subjects}

To construct \datasetns, we started by selecting representative open-source Android apps for investigation. F-Droid~\cite{fdroid} is a well known open-source Android app database, which indexes around 2,000 apps of different maturity levels. To search for suitable app subjects, we defined four criteria: (1) the app should have more than 10,000 downloads on the market (\textit{popular}), (2) the app should have a public issue tracking system (\textit{traceable}), (3) the app's code repository should contain over 100 code revisions (\textit{actively-maintained}), and (4) the app should contain at least 1,000 lines of Java code (\textit{large-scale}). With such constraints, we obtained a list of 34 subjects from F-Droid, most of which are hosted on GitHub~\cite{github}. Table~\ref{tab:subjects} provides their basic information, including (1) the app name, (2) category, (3) user rating on the Google Play store (5.0 is the highest rating), (4) number of downloads, (5) code size (lines of Java code), and (6) number of code revisions. As we can see, our subjects are diverse (covering 14 different app categories), large-scale (36.3 KLOC on average), popular (with millions of downloads), and well-maintained (with 3,653 code revisions on average).

\subsection{Collecting Resource Leak Bugs}
\begin{figure}
\includegraphics[width=\columnwidth]{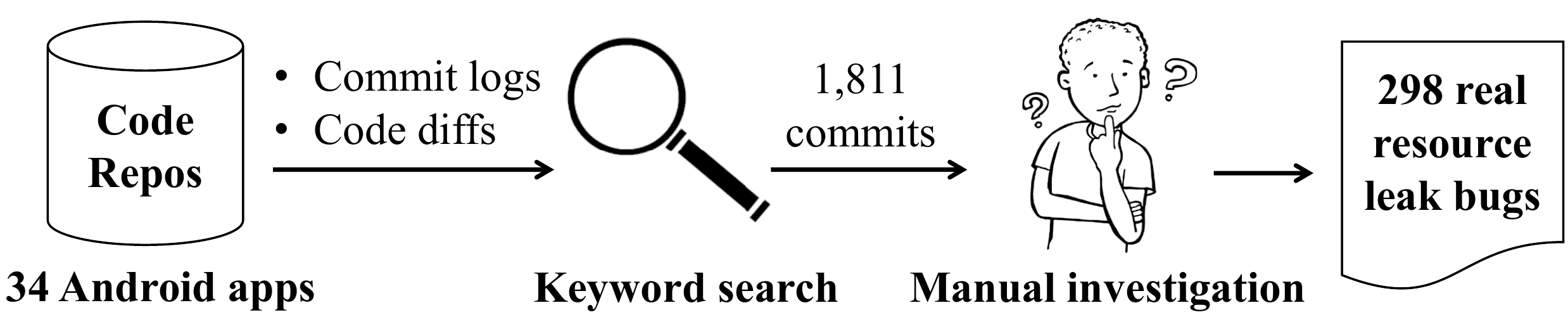}
\caption{Resource Leak Bug Collection Process}
\label{fig:bug_col}
\end{figure}

\newcolumntype{P}[1]{>{\centering\arraybackslash}p{#1}}
\begin{table}
	\centering
	\caption{Keywords for Mining Commit Logs}
	\label{tab:log_keywords}
	\renewcommand{\arraystretch}{1.1}
	\renewcommand{\tabcolsep}{0pt}
	\begin{tabular}{|P{0.2\columnwidth}|P{0.2\columnwidth}|P{0.2\columnwidth}|P{0.2\columnwidth}|P{0.2\columnwidth}|}
		\hline
		leak & leakage & release & recycle & cancel \\ \hline
		unload & unlock & unmount & unregister & close \\ \hline 
	\end{tabular}
\end{table}

\begin{table}
	\centering
	\caption{Keywords for Mining Code Diffs}
	\label{tab:diff_keywords}
	\renewcommand{\arraystretch}{1.1} %
	\renewcommand{\tabcolsep}{0pt}
	\begin{tabular}{|P{0.22\columnwidth}|P{0.36\columnwidth}|P{0.42\columnwidth}|}
		\hline
		\texttt{\scriptsize .close(} & \texttt{\scriptsize .release(} & \texttt{\scriptsize .removeUpdates(} \\ \hline
		\texttt{\scriptsize .unlock(} & \texttt{\scriptsize .stop(} & \texttt{\scriptsize .abandonAudioFocus(} \\ \hline 
		\texttt{\scriptsize .cancel(} & \texttt{\scriptsize .disableNetwork(} & \texttt{\scriptsize .stopPreview(} \\ \hline  
		\multicolumn{2}{|c|}{\texttt{\scriptsize .stopFaceDetection(}} & \texttt{\scriptsize .unregisterListener(} \\ \hline
	\end{tabular}
\end{table}

To collect fixed resource leak bugs, we mined the code repository of the 34 app subjects. Figure~\ref{fig:bug_col} illustrates the overall process, which is semi-automated and contains two major steps: (1) keyword search and (2) manual investigation.

\textbf{Keyword search.} The purpose of keyword search is to find interesting code revisions (or commits) that contain fixes to resource leak bugs. A code repository may contain a large number of code revisions. When committing each code revision, developers usually provide a natural language message to summarize their changes, a.k.a. the \textit{commit log}. The version control systems (e.g., Git) can help compute and visualize the differences between the committed version and the last historical version, a.k.a. the \textit{code diff}. Since developers may mention that they fixed certain resource leaks in commit logs and such fixes need to add code that invoke designated APIs to release resources, we defined two sets of keywords to search for interesting commit logs and code diffs, respectively. The keywords are listed in Table~\ref{tab:log_keywords} and Table~\ref{tab:diff_keywords}. The keywords in Table~\ref{tab:diff_keywords} are formulated from the state-of-art work for Android resource leak detection~\cite{relda2}, which provides a comprehensive list of resource acquiring and releasing APIs. The keywords in Table~\ref{tab:log_keywords} are general natural language words related to resource management.\footnote{While we do not claim the completeness of our keywords, our approach already successfully located a large number of real resource leak bugs in the code repositories of 32 of our 34 app subjects.} Such keywords are also needed due to two reasons. First, there is no guarantee that the set of resource releasing APIs provided by existing work is complete. Second, developers may wrap the resource releasing API calls in self-defined methods and invoking them to release resources.

To search for interesting commit logs, we first transformed all commit logs into a canonical form: containing only lower case letters and no punctuation marks. We then removed certain patterns of phrases, which accidentally include our keywords but are irrelevant, from each commit log. For instance, we removed the phrases that match these two regular expressions: ``\texttt{\small release (v$\vert$ver)?[0-9]+}'' and ``\texttt{\small close issue \#?[0-9]+}'' as phrases such as ``release v1.0.1'' and ``close issue \#168'' frequently occur in commit logs.\footnote{In our mining scripts, we defined 32 removal patterns after randomly sampling 1,000+ commit logs. We skip the details in this paper.} Next, we split each processed commit log into a vector of words and stemmed each word into its root form. Stemming~\cite{stemming} is necessary because the natural language words may be in different forms. For example, the verb ``release'' may be in its gerund form ``releasing'' in certain commit logs and we need to stem it into its root from ``releas''. After stemming, we applied the stemmed form of the keywords in Table~\ref{tab:log_keywords} for searching.

To search for interesting code diffs, we looked for those diffs that contains lines (1) starting with the ``+'' symbol (indicating code addition as fixing resource leak bugs requires developers to add API calls to release the concerned resources) and (2) containing a keyword from Table~\ref{tab:diff_keywords} (for matching API signatures). With the above two searching steps, we obtained a set of code revisions that contain either interesting commit logs or interesting code diffs. Column 7 of Table~\ref{tab:subjects} lists the number of such code revisions we found for each app subject.

\textbf{Manual investigation.} In total, keyword search located 1,811 interesting code revisions. We then carefully investigated each of them to check whether it fixes resource leaks or not. The process took us several months and in the end we successfully found \numbugs resource leak bugs from 176 code revisions (some code revisions fix multiple resource leaks). The remaining code revisions are irrelevant but retrieved because their commit logs accidentally contain our search keywords or their code diffs contain the addition of resource releasing API calls for other purposes (e.g., refactoring). The last column of Table~\ref{tab:subjects} lists the number of real resource leak bugs we found for each app subject. As we can see, \textit{32 of the 34 apps (94.1\%) were infected by resource leaks, which is a scary fact suggesting the pervasiveness of resource leak bugs in real-world Android apps.} Then for each bug, we further collected the following information to construct \datasetns: (1) the buggy code, (2) the bug-fixing patch, and (3) the bug report if we can find it in the app's issue tracking system. These data will be used in our empirical study later.

\section{Empirical Study}\label{sec:empirical}
To answer RQ1--3, we carefully studied each bug in \dataset and examined the relevant code (e.g., patches) and data (e.g., bug reports). This section reports our observations.

\subsection{RQ1: Resouce Type and Consequence of Leak}
\newcolumntype{M}[1]{>{\centering\arraybackslash}m{#1}}
\begin{table}[t!]
	\begin{threeparttable}
	\centering
	\caption{Resource Leak Bug Statistics}
	\label{tab:stats}
	\renewcommand{\tabcolsep}{0pt}
	\renewcommand{\arraystretch}{1.1}
	\begin{tabular}{|P{0.05\columnwidth}|P{0.85\columnwidth}|P{0.1\columnwidth}|}
		\hline
		& \textbf{Concerned Java class (consequence of leak)} & \textbf{\# bugs} \\ \hline
		\multirow{11}{*}{\rotatebox[origin=c]{90}{Android platform resources}} & \texttt{\scriptsize android.database.Cursor (I)} & 144 \\ \cline{2-3} 
		& \texttt{\scriptsize android.database.sqlite.SQLiteDatabase (I)} & 13 \\ \cline{2-3} 
		& \texttt{\scriptsize android.os.PowerManager.WakeLock (II)} & 9 \\ \cline{2-3} 
		& \texttt{\scriptsize android.media.MediaPlayer (I)} & 5 \\ \cline{2-3} 
		& \texttt{\scriptsize android.net.wifi.WifiManager.WifiLock (II)} & 2 \\ \cline{2-3} 
		& \texttt{\scriptsize android.location.LocationListener (II)} & 2 \\ \cline{2-3}
		& \texttt{\scriptsize android.database.sqlite.SQLiteOpenHelper (I)} & 1 \\ \cline{2-3} 
		& \texttt{\scriptsize android.view.MotionEvent (I)} & 1 \\ \cline{2-3}
		& \texttt{\scriptsize android.os.ParcelFileDescriptor (I)} & 1 \\ \cline{2-3}
		& \texttt{\scriptsize android.os.Parcel (I)} & 1 \\ \cline{2-3}
		& \texttt{\scriptsize android.hardware.Camera (III)} & 1 \\ \hline
		\multirow{26}{*}{\rotatebox[origin=c]{90}{General Java platform resources}} & \texttt{\scriptsize java.io.InputStream (I)} & 32 \\ \cline{2-3} 
		& \texttt{\scriptsize java.io.FileInputStream (I)} & 12 \\ \cline{2-3} 
		& \texttt{\scriptsize java.io.FileOutputStream (I)} & 10 \\ \cline{2-3} 
		& \texttt{\scriptsize java.io.BufferedReader (I)} & 9 \\ \cline{2-3} 
		& \texttt{\scriptsize java.io.FilterOutputStream (I)} & 9 \\ \cline{2-3} 
		& \texttt{\scriptsize java.io.OutputStream (I)} & 7 \\ \cline{2-3} 
		& \texttt{\scriptsize java.io.FilterInputStream (I)} & 4 \\ \cline{2-3} 
		& \texttt{\scriptsize org.apache.http.impl.client.DefaultHttpClient (I)} & 4 \\ \cline{2-3} 
		& \texttt{\scriptsize java.io.BufferedOutputStream (I)} & 3 \\ \cline{2-3} 
		& \texttt{\scriptsize java.util.concurrent.Semaphore (III)} & 3 \\ \cline{2-3} 
		& \texttt{\scriptsize java.io.BufferedWriter (I)} & 2 \\ \cline{2-3} 
		& \texttt{\scriptsize java.io.ByteArrayOutputStream (I)} & 2 \\ \cline{2-3} 
		& \texttt{\scriptsize java.io.OutputStreamWriter (I)} & 2 \\ \cline{2-3} 
		& \texttt{\scriptsize java.net.Socket (I)} & 2 \\ \cline{2-3}
		& \texttt{\scriptsize java.util.Scanner (I)} & 2 \\ \cline{2-3} 
		& \texttt{\scriptsize org.apache.http.impl.client.HttpClient (I)} & 2 \\ \cline{2-3}
		& \texttt{\scriptsize java.io.ObjectInputStream (I)} & 2 \\ \cline{2-3}
		& \texttt{\scriptsize java.io.ObjectOutputStream (I)} & 2 \\ \cline{2-3}
		& \texttt{\scriptsize java.io.PipedOutputStream (I)} & 2 \\ \cline{2-3}
		& \texttt{\scriptsize com.fasterxml.jackson.core.JsonParser (I)} & 1 \\ \cline{2-3} 
		& \texttt{\scriptsize com.google.gson.JsonParser (I)} & 1 \\ \cline{2-3} 
		& \texttt{\scriptsize java.io.DataOutputStream (I)} & 1 \\ \cline{2-3}
		& \texttt{\scriptsize java.io.InputStreamReader (I)} & 1 \\ \cline{2-3}
		& \texttt{\scriptsize java.io.PipedInputStream (I)} & 1 \\ \cline{2-3}
		& \texttt{\scriptsize java.util.Formatter (I)} & 1 \\ \cline{2-3}
		& \texttt{\scriptsize java.util.logging.FileHandler (I)} & 1 \\ \hline
	\end{tabular}
\end{threeparttable}
\end{table}

\begin{figure}[ht!]
	\centering
	\includegraphics[width=0.9\columnwidth]{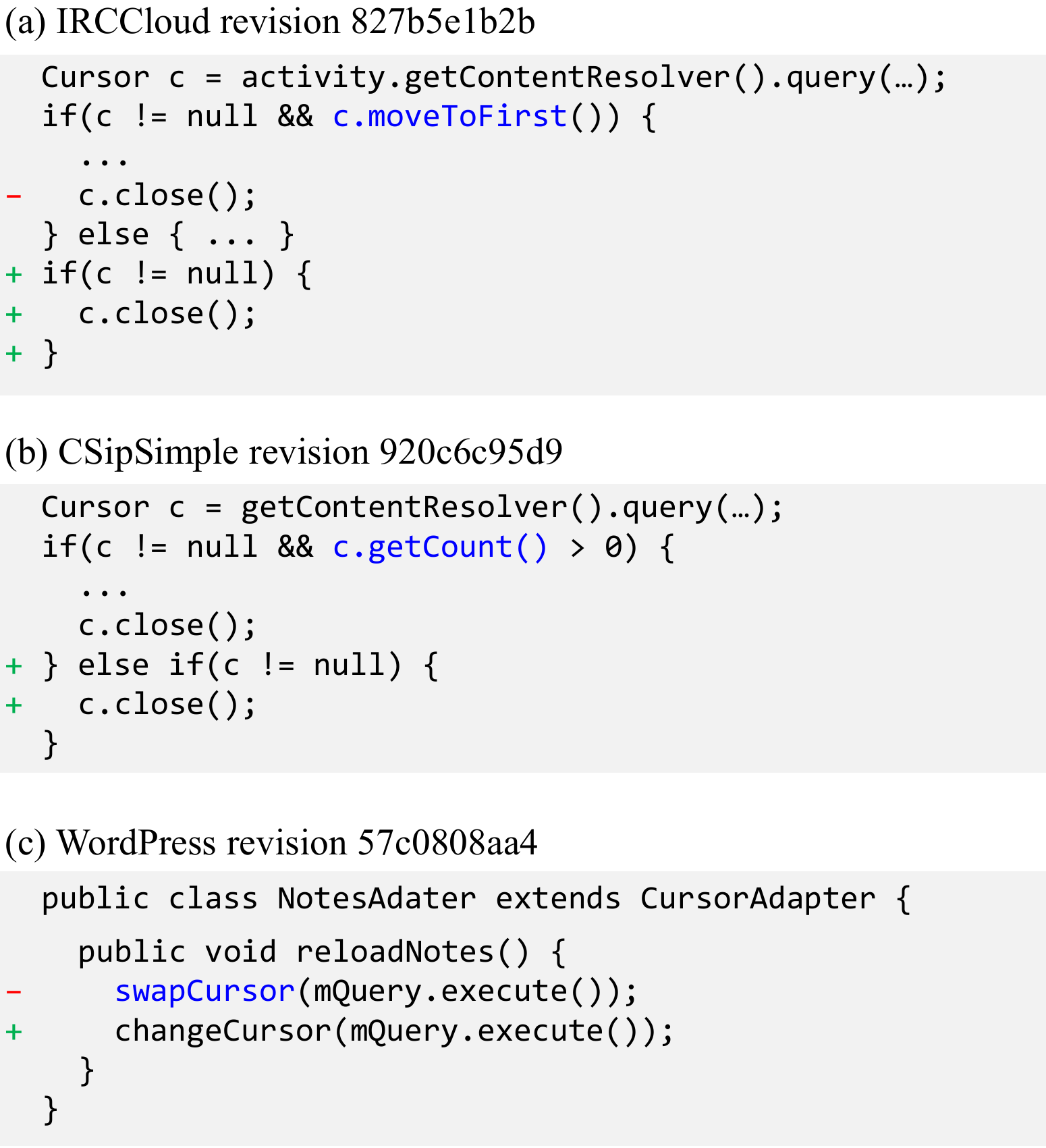}
	\caption{Leaks of Database Cursors Due to API Misuses}
	\label{fig:api_misuse}
\end{figure}

\textbf{Resource types}. Overall, the \numbugs bugs in \dataset cover 37 different resource classes listed in Table~\ref{tab:stats}. As we can see from the table, 60.4\% of the bugs (180 of \numbugsns) concern resources that are specific to Android platforms. For instance, the SQLite database is widely-used in Android apps and we found 144 bugs in \dataset leaking SQLite database cursors (see Figure~\ref{fig:api_misuse} for examples). The rest 118 bugs (39.6\%) leak general Java platform resources, of which I/O streams account for the majority. It is not surprising that the percentage of Java platform resource leaks is quite high since Android apps are essentially Java programs and can use various Java libraries to level system resources for computational purposes.

\textbf{Consequence of resource leaks.} Resource leaks are generally considered as non-functional issues that do not cause immediate fail-stop consequences such as app crashes. To understand the consequences of the bugs in \datasetns, we studied the bug reports and API documentations~\cite{api}. We observed three types of major consequences.

\begin{itemize}[leftmargin=*,itemsep=2pt]
	\item{Most of the resource leak bugs (281 of \numbugsns, marked with ``I'' in Table~\ref{tab:stats}) will mainly lead to memory waste, which can gradually slow down the whole system. For example, an \texttt{\small android.database.Cursor} object has native file handles behind it because SQLite database uses indexed files for providing query functions~\cite{api}. Forgetting to close a cursor object will prevent its associated file handles from being released and thus causing memory waste. In worst cases, an app could crash due to out of memory exceptions.}
	
	\item{The second common consequence is energy waste, concerning 13 bugs in \dataset (marked with ``II'' in Table~\ref{tab:stats}). These bugs leak wake lock, Wi-Fi lock, and sensor-related resources, which are specific to Android platforms. For example, wake locks provide a mechanism to indicate that an app needs the device to stay awake for long-running operations (e.g., large file downloading). Calling the \texttt{\small acquire()} API on an \texttt{\small android.os.PowerManager.WakeLock} instance will force the device to stay awake. While this interface is convenient, the Android API guides~\cite{api} also warn developers to carefully use wake locks: \textit{``Call release() when you are done and don't need the lock anymore. It is very important to do this as soon as possible to avoid running down the device's battery excessively''}. Nonetheless, developers still often make mistakes and we found nine wake lock leaks in \datasetns. Besides wake locks, \dataset also contains bugs that forget to release Wi-Fi locks, which are used to keep the Wi-Fi radio on for network communications, and sensor listeners, which are registered to obtain continuous updates from phone sensors. Such bugs can also lead to serious energy waste.}
	
	\item{The remaining four bugs (marked with ``III'' in Table~\ref{tab:stats}) concern exclusive resources: camera and semaphore (for restricting the number of concurrent threads). Forgetting to release them can affect app functionalities. For example, Android API guides ask developers to release cameras when their apps finish using them to avoid affecting their own or other apps: \textit{``If your application does not properly release the camera, all subsequent attempts to access the camera, including those by your own application, will fail and may cause your or other applications to be shut down''.} Similarly, forgetting to release semaphores can block acquiring threads and the corresponding computational tasks, leading to unexpected app behaviors or even crashes in case apps stop responding to user interactions~\cite{anr}.}
	
\end{itemize}

\vspace{10pt}
\noindent\fbox{
	\parbox{0.95\linewidth}{
		\textbf{Answer to RQ1:} \textit{\dataset features a diverse set of resource leak bugs covering 37 different resource classes. Most bugs leak Android-specific resources and can waste memory, drain battery power, or even crash an app.} 
	}
}
\vspace{2pt}

\subsection{RQ2: Resource Leak Extent}\label{ssec:leak_extent}

\begin{figure}
	\centering
	\includegraphics[width=0.85\columnwidth]{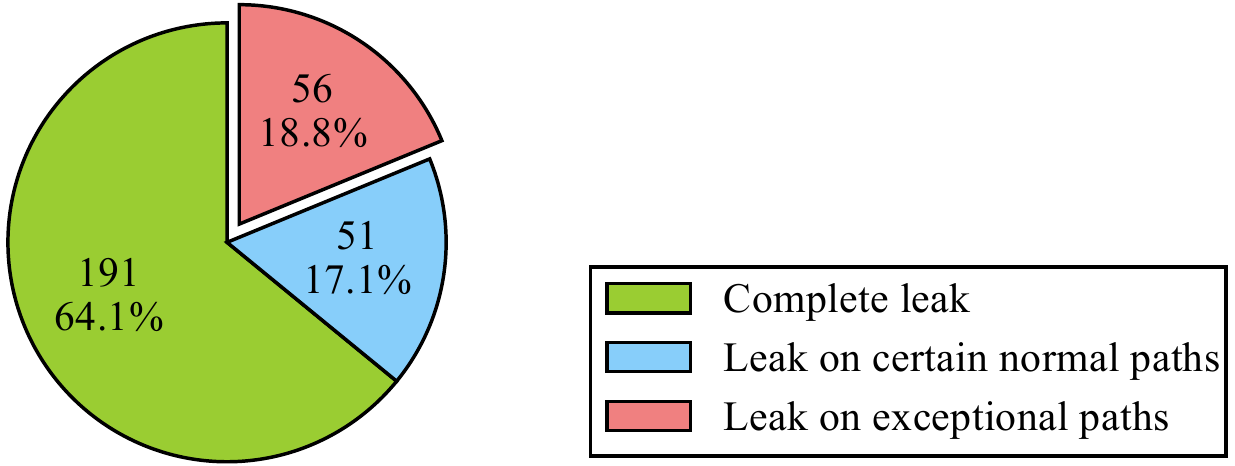}
	\caption{Resource Leak Extent}
	\label{fig:extent}
\end{figure}

Resource leaks are code omission faults, where developers forget to release used system resources on certain program execution paths. Depending on what execution paths the resources are leaked on, we categorize the bugs in \dataset into three categories:

\begin{itemize}[leftmargin=*, itemsep=2pt]
	\item{\textit{Complete leak:} developers completely forget to release the system resources after use.}
	
	\item{\textit{Leak on exceptional paths:} the system resources are properly released in normal execution paths, but fail to be released in case of runtime exceptions.}
	
	\item{\textit{Leak on certain normal paths:} the system resources are released on some program paths during normal executions (i.e., without runtime exceptions), but fail to be released on the others. Note that this category includes resource leaks that occur under app-specific erroneous conditions, where no runtime exceptions are thrown or handled but the concerned app enters a wrong internal state.\footnote{We only observed a couple of such cases in \dataset and therefore do not specifically discuss them in this paper.}}
\end{itemize} 

The categorization is manually performed by studying the bug patching code. Figure~\ref{fig:extent} presents the result. The majority (64.1\%) bugs in \dataset completely leak system resources on all program execution paths, while the remaining 35.9\% partially leak system resources on certain normal or exceptional paths. According to the existing studies, it is understandable that Java developers, even experienced ones, can easily fail to release all system resources along all possible exceptional paths~\cite{tracker}. However, in \datasetns, we observed that resource leaks on exceptional paths only account for a minority (18.8\%). Instead, leaks during normal executions are the majority (81.2\% = 64.1\% + 17.1\%), which is unexpected. So we further analyzed the bugs in \dataset to figure out why resource leaks are so common on normal program execution paths and found several major reasons.

\begin{figure}
	\centering
	\includegraphics[width=0.9\columnwidth]{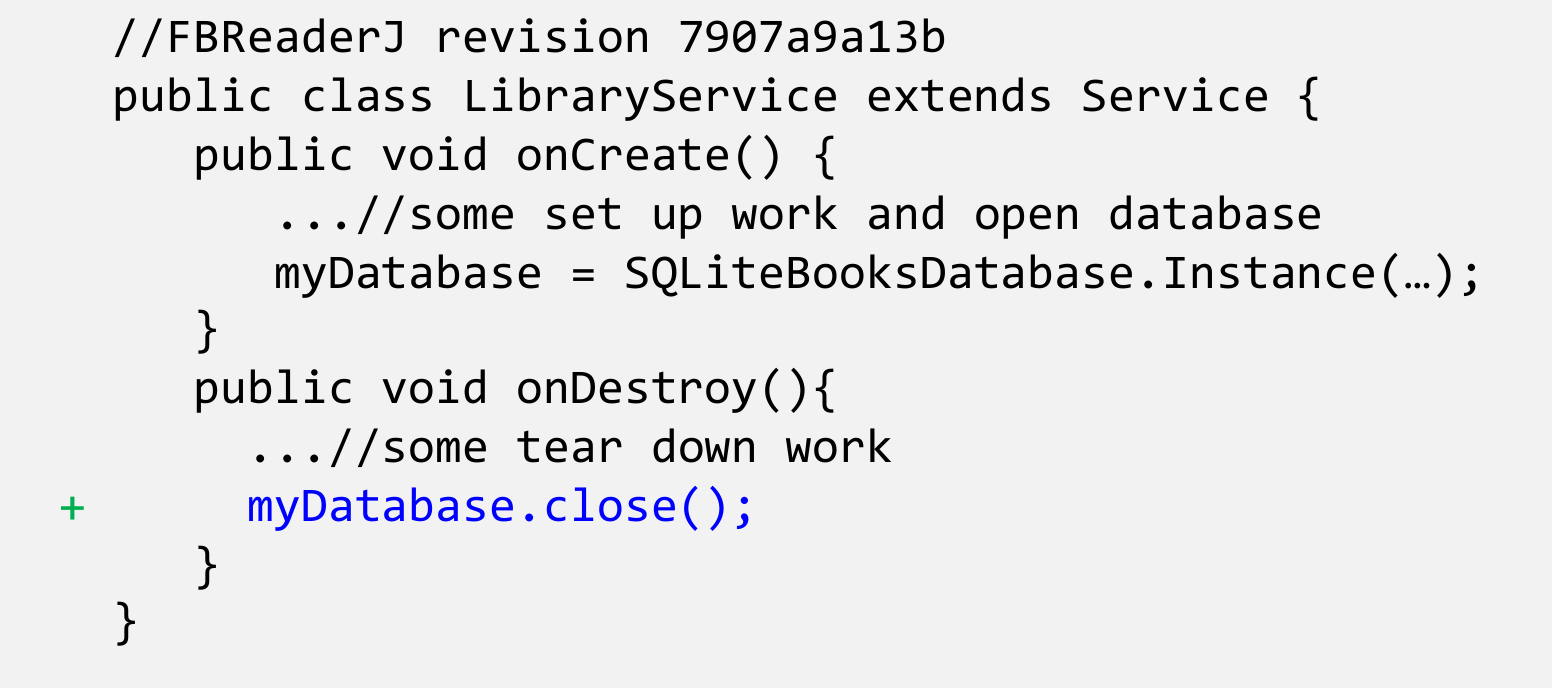}
	\caption{A Resource Leak Bug Due to Complex Component Lifecycle}
	\label{fig:lifecycle}
\end{figure}

\begin{figure}
	\centering
	\includegraphics[width=0.9\columnwidth]{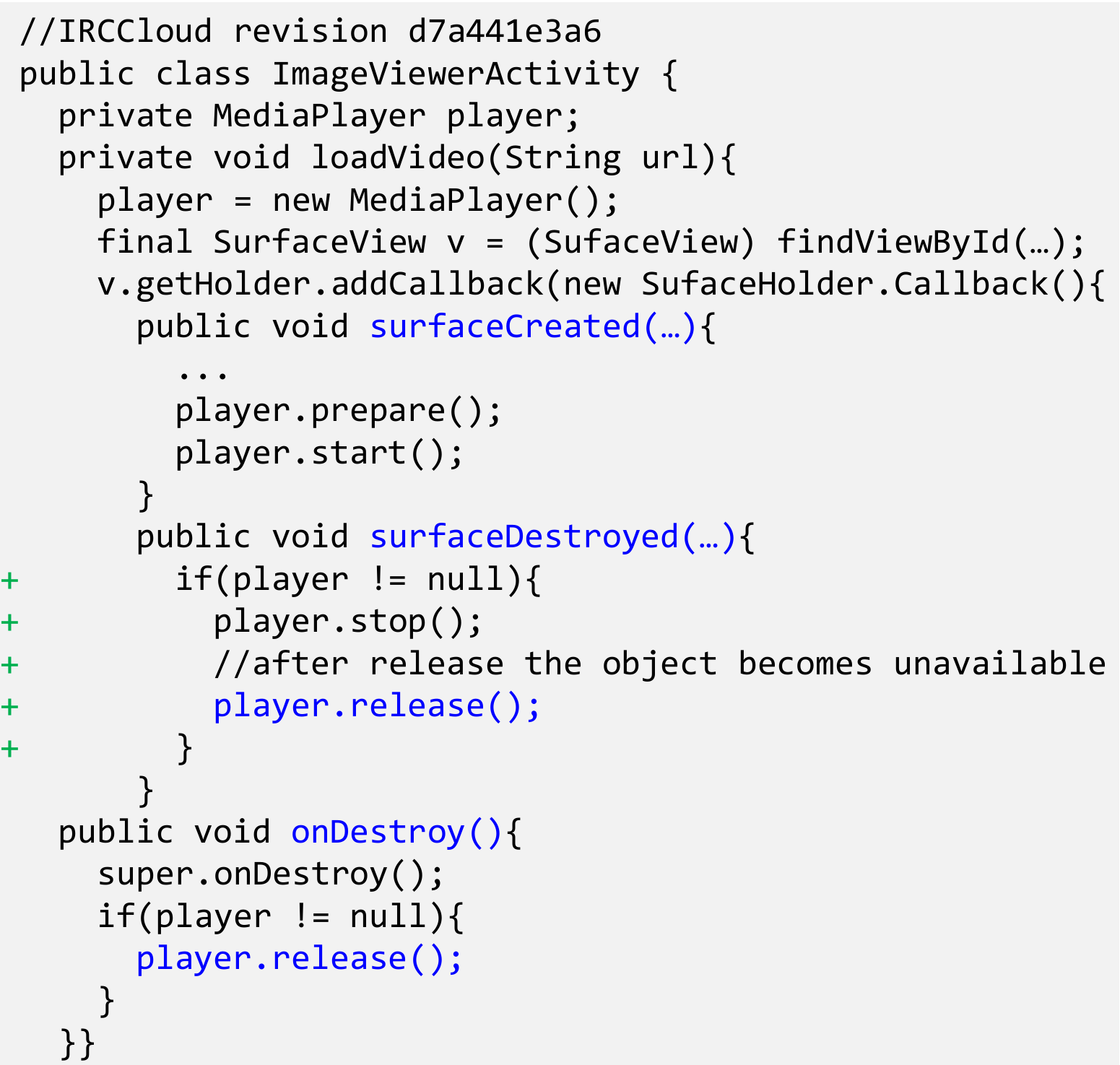}
	\caption{A Resource Leak Bug Due to Unexpected User Interactions}
	\label{fig:user_interaction}
\end{figure}

\begin{figure}[t!]
	\centering
	\includegraphics[width=0.9\columnwidth]{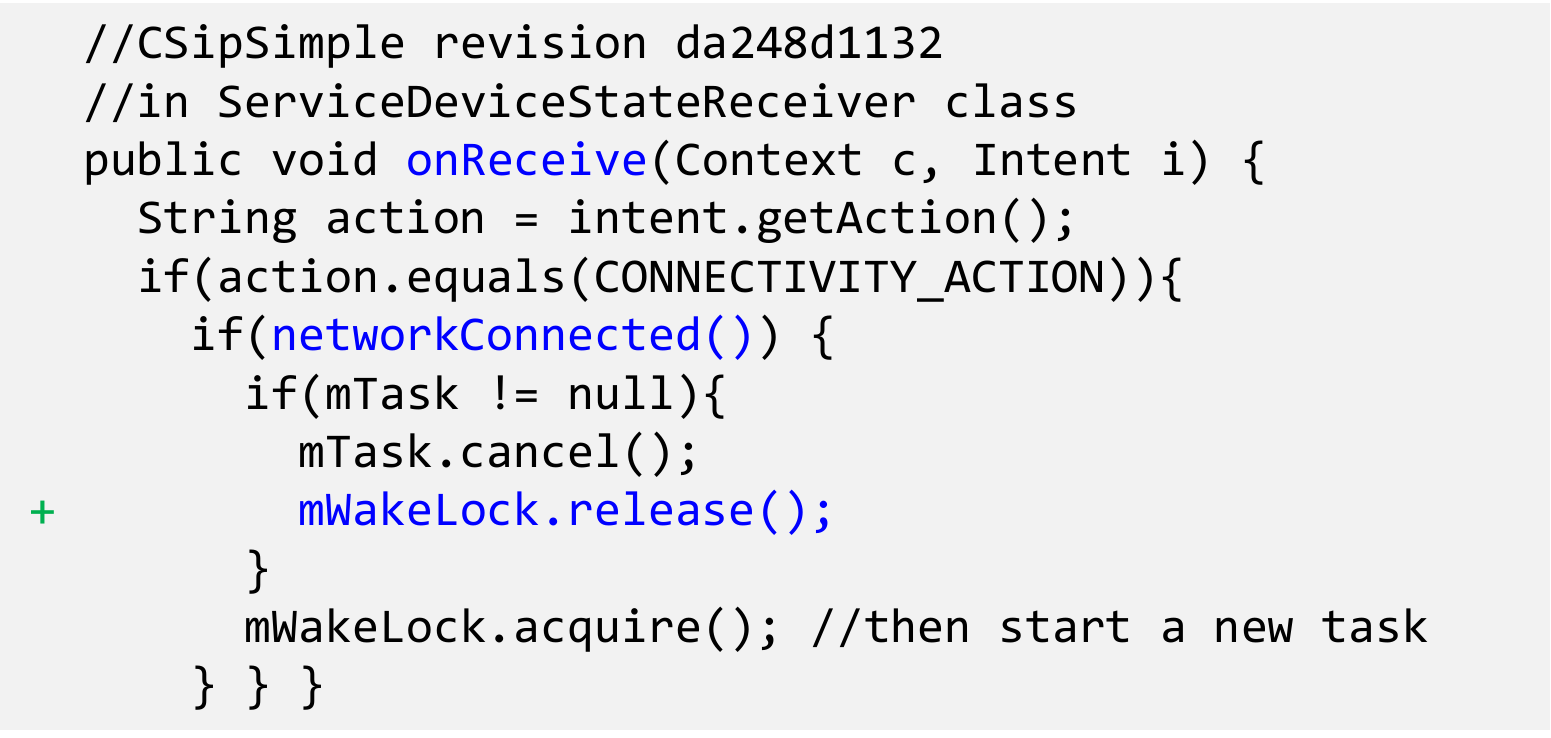}
	\caption{A Resource Leak Bug Due to Environment Interplay}
	\label{fig:network}
\end{figure}

\begin{figure}
	\centering
	\includegraphics[width=0.9\columnwidth]{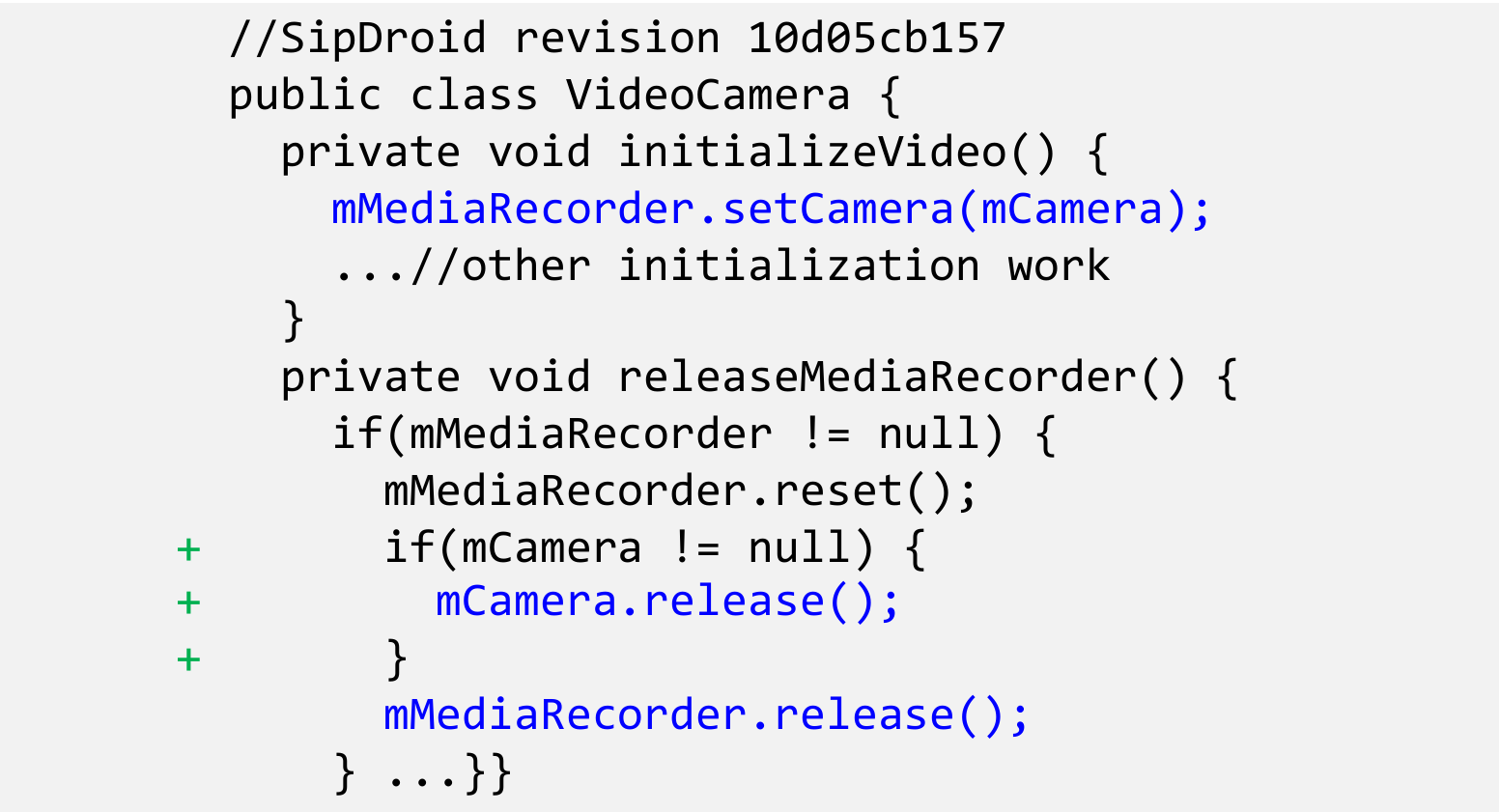}
	\caption{A Resource Leak Bug Due to API Unfamilarity}
	\label{fig:camera}
\end{figure}

\begin{figure}[htbp!]
	\centering
	\includegraphics[width=0.9\columnwidth]{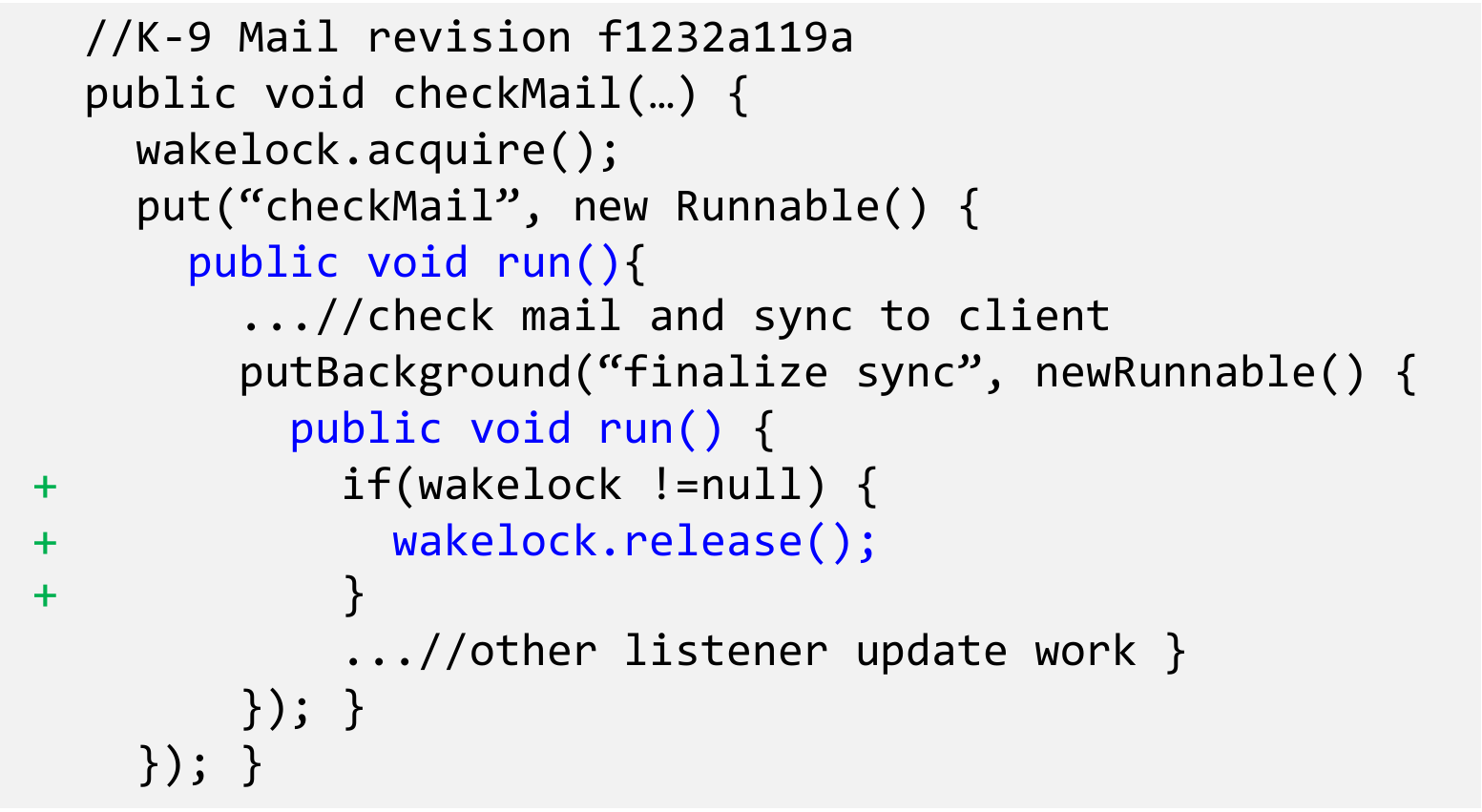}
	\caption{A Resource Leak Bug Due to Concurrency}
	\label{fig:thread}
\end{figure}

\begin{itemize}[leftmargin=*, itemsep=4pt]
	\item{\textit{Complex app component lifecycle.} As we mentioned in Section~\ref{sec:bg}, Android apps are comprised of different app components. Each app component is required to follow a prescribed lifecycle that defines how this component is created, used, and finally destroyed~\cite{api}. At runtime, the lifecycle event handlers (i.e., callbacks) defined in an app component will be invoked by the Android OS when the component enters certain lifecycle stages. For instance, when a background service component is started, its \texttt{\small onCreate()} handler (see Figure~\ref{fig:lifecycle} for example)\footnote{The code in all figures in this paper has been simplified for readability.\vspace{8pt}} will be invoked. When the service finishes its allocated computational task, it will be destroyed and the \texttt{\small onDestroy()} handler will be invoked. Typically, when starting a service component, certain system resources need to be acquired for later computation (e.g., database connection in the FBReaderJ~\cite{fbreader}, an E-book reading app, as shown in Figure~\ref{fig:lifecycle}). The acquired resources should be released when the service is destroyed. However, due to the complexity that the resources are acquired and released in different lifecycle stages (i.e., in different callback methods), developers often forget to release the resource properly such as in the FBReaderJ example. We observed tens of such bugs in \dataset and some involve acitivity components with multiple fragments~\cite{fragments}, whose lifecycle is much more complicated than services.}
		
	\item{\textit{Unexpected user interactions.} Android apps are event-driven. When interacting with users, an app may receive all kinds of inputs and handle various events. It is difficult for developers to anticipate all possible user interactions when building apps. Therefore, it is common that developers forget to manage resources correctly in certain unexpected scenarios. Figure~\ref{fig:user_interaction} provides such an example. The app IRCCloud, a group chatting app, contains an activity class \texttt{\small ImageViewerActivity} to view images and videos. The activity class uses a surface view to implement a floating window for playing videos. In normal cases, when the floating window pops up, the underlying media player is started (see the \texttt{\small surfaceCreate()} callback). When users quit the activity after watching the video (e.g., switching to another screen), the media player is released (see the activity's \texttt{\small onDestroy()} handler). However, if the users stay at the activity screen to watch other videos after they finish one video (there will be a new floating window poping up and a new media player object constructed), the floating window for playing the finished video will lose focus and the associated media player will never be released. This mistake happened because developers did not carefully think about possible user interactions to avoid resource leaks. We also see many other similar mistakes in \dataset such as leaking I/O streams when users cancel file downloading.}
		
	\item{\textit{Environment interplay.} Besides handling user inputs, many Android apps also need to frequently handle environmental changes (e.g., changes of user location) to provide context-aware services. Similar to user inputs, environmental conditions are hard to predict. It is common that developers make resource management mistakes when handling environmental changes. Figure~\ref{fig:network} gives an example bug. The app CSipSimple~\cite{csipsimple}, an Internet calling app, uses a reference counted wake lock to keep device awake for phone calls. It monitors the network status by the \texttt{\small ServiceDeviceStateReceiver} broadcast receiver, which actively listens to system-wide broadcast messages (e.g., network state changes). Whenever the network connection is restored, CSipSimple acquires a wake lock for performing computational tasks. Under stable network conditions, there is usually just one acquisition of wake lock and everything will work smoothly. However, when the network condition is poor (i.e., frequent disconnection and reconnection), the code for acquiring the wake lock will be executed many times, leading to the repetitive acquisition of held wake locks. The consequence is that this reference-counted wake lock will be not properly closed with one call to the releasing API and the device will stay awake indefinitely, causing huge energy waste. Later, developers fixed the leak by releasing the held wake lock (the lock is held when there is a long running task \texttt{\small mTask}) before acquiring a new one (in revision \texttt{\small da248d1132}~\cite{csipsimple}).}
	
	\item{\textit{API unfamiliarity.} Android is a relatively new platform, which exposes over ten thousand public APIs~\cite{permission} to ease app development. In practice, it is generally impossible for developers to get familiar with the specification of each Android API before developing apps. Therefore, they often make mistakes when using unfamiliar APIs and resource leaks can arise in such cases. We discuss a typical example in Figure~\ref{fig:camera}. The app SipDroid~\cite{sipdroid}, a popular Internet calling app similar to CSipSimple, uses phone cameras for video recording. The developers were aware of the importance of correct resource management and released the \texttt{\small mMediaRecorder} object after use. However, they mistakenly thought that the enclosing \texttt{\small mCamera} object would be transitively released (i.e., like in Java, calling the \texttt{\small close()} method on a \texttt{\small BufferedReader} object will also close the enclosing input streams). This resulted in a serious resource leak, which can stop other apps that request the use of camera from functioning, as camera is an exclusive resource. The developers later realized the mistake and fixed it by releasing the camera as well (in revision \texttt{\small 10d05cb157}~\cite{sipdroid}).}
	
	\item{\textit{High level of concurrency.} Lastly, Android apps adopt a single thread model: all app components that run in the same process are instantiated in the app's main thread (a.k.a. UI thread), which is created by the Android OS when the app is launched, and system calls to each component are dispatched from that thread~\cite{process}. Hence, Android apps usually leverage various concurrent programming constructs~\cite{async} (e.g., \texttt{\small Android.os.AsyncTask} and \texttt{\small java.lang.Thread}) to perform intensive work (e.g., network communications, database queries) in other threads in order not to block the main thread, which would lead to poor runtime performance or even app not responding (ANR) errors~\cite{anr}. Such a high level of concurrency can easily cause resource leaks, especially when the resource acquiring and releasing operations are not in the same thread. Figure~\ref{fig:thread} provides a typical example. The app K-9 Mail, a widely-used email client, acquires a wake lock for keeping device awake to check emails. For syncing emails to the local folders, it creates a worker thread to communicate with the server. When the syncing is finished, the worker thread further creates another thread to notify listeners. The acquired wake lock should be released after mail syncing. However, developers forgot this as there are multiple threads involved and the wake lock acquiring operation is quite far away in the code that runs in the app's main thread (in Figure~\ref{fig:thread}, the wake lock operations are close to each other because the code was simplified to ease understanding). They later figured out the mistake and released the wake lock properly.}
\end{itemize}

The root causes of the prevalence of resource leaks in Android apps also pose challenges for bug detection. For dynamic analyses, how to effectively generate user interactions and simulate environment interplays to trigger the resource leaking scenario is a difficult task. For static analyses, how to handle the implicit control flows among various callbacks for inferring possible execution paths is a major challenge. Besides, static analyses also need to precisely model concurrency and perform points-to analysis, since many resources are not acquired and released in the same method or thread.)

\vspace{10pt}
\noindent\fbox{
	\parbox{0.95\linewidth}{
		\textbf{Answer to RQ2:} \textit{The majority of bugs in \dataset completely leak system resources on all program execution paths. Complex app component lifecycle, unexpected user interactions, environment interplay, API unfamiliarity, and high level of concurrency are five major reasons why Android apps often leak resources during normal executions.} 
	}
}
\vspace{2pt}

\begin{figure}[ht!]
	\centering
	\includegraphics[width=0.9\columnwidth]{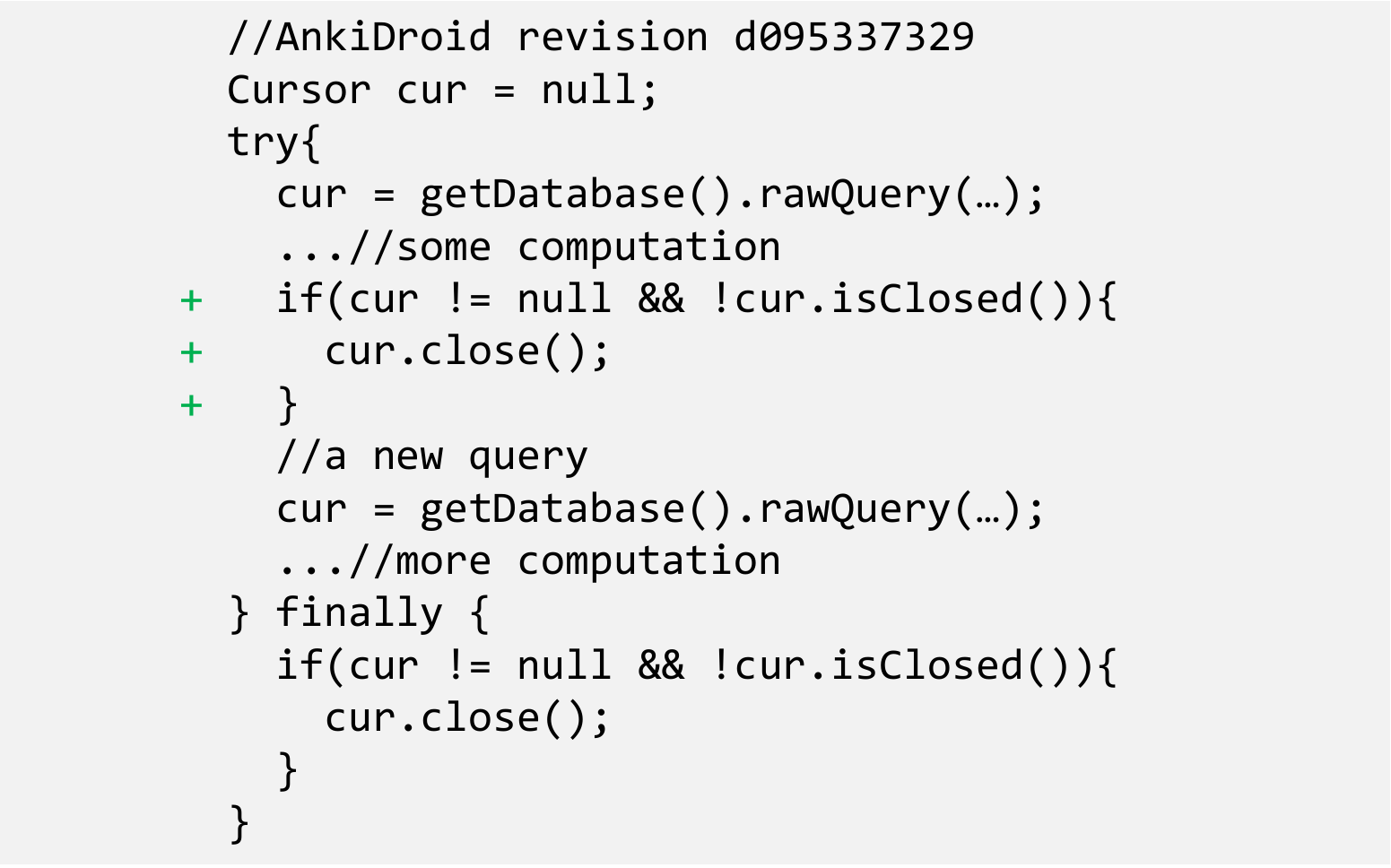}
	\caption{Resource Leak Due to Losing Resource Object Reference}
	\label{fig:lost_ref}
\end{figure}

\begin{figure}[ht!]
	\centering
	\includegraphics[width=0.9\columnwidth]{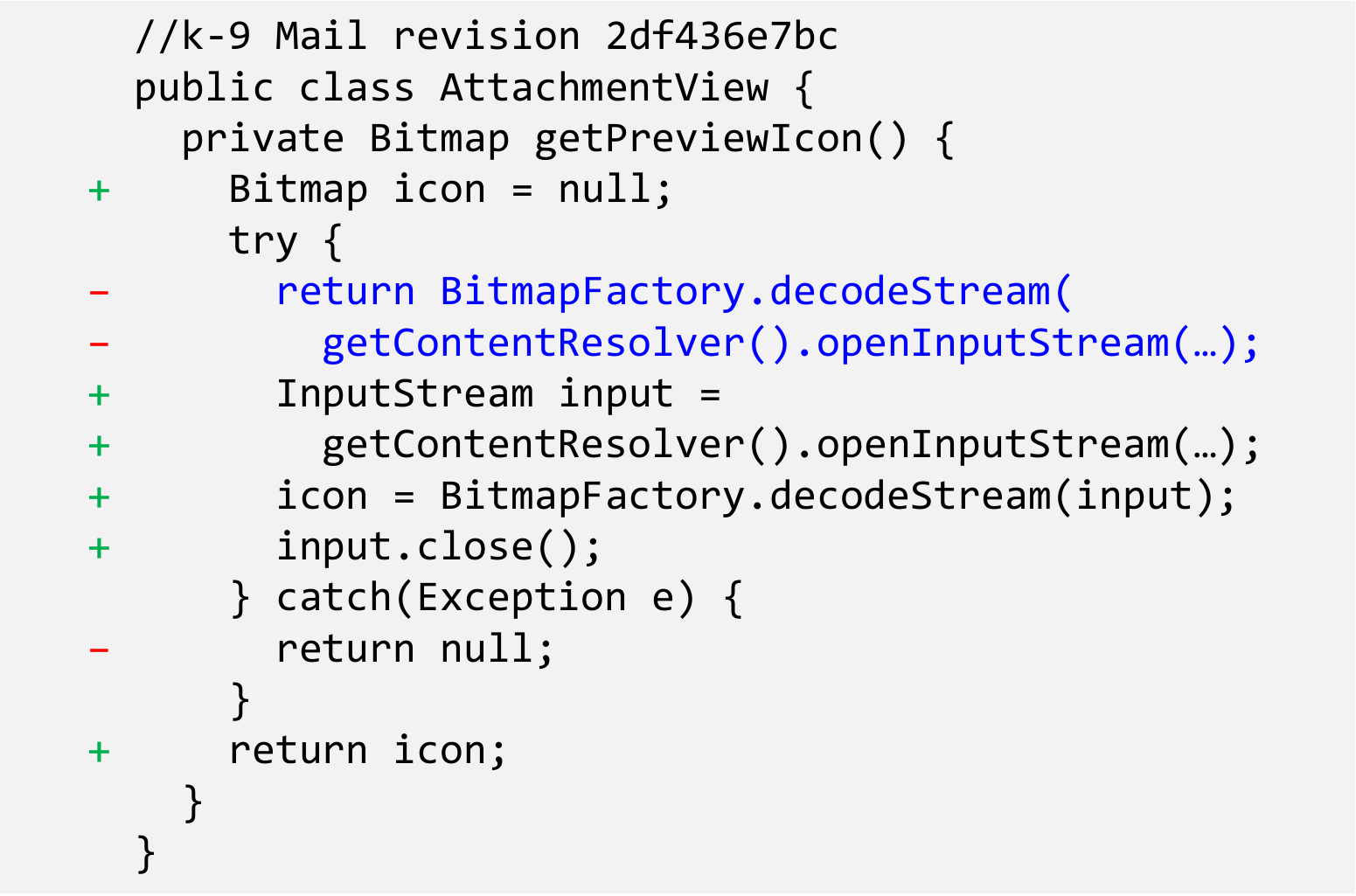}
	\caption{Resource Leak Due to the Lack of Resource Object Reference}
	\label{fig:no_ref}
\end{figure}

\subsection{RQ3: Common Fault Patterns}
To answer RQ3, we analyzed all bugs in \dataset and tried to figure out the mistakes made by developers. We observed that in most cases, developers simply forgot the releasing of system resources after use due to the complexity of their programs or unawareness of the necessity to release resources (we discuss several reasons why they made such mistakes in Section~\ref{ssec:leak_extent}). Nonetheless, we also observed three patterns of faults that recur across our studied apps. We discuss them with examples in the following.

\begin{itemize}[leftmargin=*,itemsep=2pt]
	\item{\textit{API misunderstanding and misuses.} As we discussed earlier, Android developers can make resource management mistakes when they use the APIs that they are not familiar with. In \datasetns, we observed three widely-used database APIs developers often misuse. The first one is the \texttt{\small moveToFirst()} API defined in the \texttt{\small android.database.Cursor} class. Calling it will move the concerned database cursor to the first row if the cursor is not empty, or return false otherwise. Many developers thought that only non-empty database cursors need to be closed (i.e., when this API returns true). Figure~\ref{fig:api_misuse}(a) gives an example bug in IRCCloud~\cite{irccloud}. The second API \texttt{\small getCount()} is also defined in the database cursor class and it returns the number of rows in a cursor. Similar to \texttt{\small moveToFirst()}, many developers use \texttt{\small getCount()} to check whether a database query returns an empty cursor and thought that only when there is at least one row returned by the query, the cursor needs to be closed. We also provide an example bug in CSipSimple~\cite{csipsimple} in Figure~\ref{fig:api_misuse}(b) for illustration. The third API is the \texttt{\small swapCursor()} API defined in the \texttt{\small android.widget.CursorAdapter} class, which is frequently used to adapt cursor data to a list view widget (a view that shows items in a vertically scrolling list). To replace the underlying database cursor associated with a \texttt{\small CursorAdapter} with a new one (e.g., after a new query), developers have two choices of APIs: \texttt{\small swapCursor(Cursor newCursor)} or \texttt{\small changeCursor(Cursor newCursor)}. The only difference between the two APIs is that the former does not close the old cursor, but returns it, while the latter closes the old cursor. In practice, developers often misuse \texttt{\small swapCursor()}, mistakenly thinking that it would close the old cursor. In Figure~\ref{fig:api_misuse}(c), we provide a bug in WordPress, an app for creating websites and blogs, and its patch for illustration.}
	
	\item{\textit{Lost reference to resource objects} is the second common pattern of faults, which would lead to resource leaks. Figure~\ref{fig:lost_ref} gives a typical example in the app AnkiDroid~\cite{ankidroid}, a popular flashcard app for education. As we can see from the code snippet, the app performs two queries consecutively to retrieve data from database. The developers were aware that database cursors need to be closed after use and put the cursor closing code in a \texttt{\small finally} block. However, since there are two queries, two cursor objects are constructed, but the local variable \texttt{\small var} only holds the reference to the second cursor object. The reference to the first cursor object was overridden after requery and therefore lost. The consequence is that the first database cursor is left unclosed, resulting in the leak of its associated resources, which would not be automatically recycled by the garbage collector~\cite{tracker}. Developers later fixed their mistake by releasing the leaked cursor (in revision \texttt{\small d095337329~\cite{ankidroid}}). Due to the prevalence of multiple queries, such faults are common in \datasetns. Besides database cursors, we also observed similar faults, where developers mistakenly override the variables that hold references to I/O streams.}
	
	\item{\textit{Lacking reference to resource objects} is the third common pattern of faults made by developers. In Android apps, resource operations are performed by invoking certain APIs on resource objects. However, in \datasetns, we observe that developers often forget to create resource object reference variables and simply put resource operations as nested method calls. Figure~\ref{fig:no_ref} gives an example in K-9 Mail, where the input stream opening API call is nested in the stream decoding API call, and there is no variable holding the reference to the underlying input stream object. Hence, developers can easily forget to release the acquired resources. Such faults affected eight apps in \datasetns.}
\end{itemize}

\vspace{10pt}
\noindent\fbox{
	\parbox{0.95\linewidth}{
		\textbf{Answer to RQ3:} \textit{We observed three common patterns of faults made by developers in \datasetns: (1) API misunderstanding and misuses, (2) lost reference to resource objects, and (3) lacking reference to resource objects.} 
	}
}

\vspace{8pt}

\section{Discussions}
\subsection{Threats to Validity}
The validity of our study results may be subject to several threats. The first is the representativeness of our selected Android app subjects. To minimize the threat, we randomly selected 34 large-scale and popular open-source Android apps from the F-Droid database~\cite{fdroid}. These apps are diverse and cover 13 different app categories. We believe that our empirical findings can generalize to a wide range of real-world Android apps. The second threat is the potential misses of real resource leak bugs in our keyword search process. To reduce the threat, we leveraged the resource operations identified by the state-of-the-art work to formulate keywords. We also used general keywords such as ``leak'', ``release'' and ``close'', aiming to maximally cover resource leak bugs that affected our app subjects. The strategy indeed helped us find a large number of fixed resource leak bugs in 33 out of our 34 app subjects and we believe they have covered most real bugs. The last threat is the errors in our manual investigation of bugs. To minimize the threat, we co-authors carefully cross-validated the results. We will also release them for public access~\cite{leakbench}.

\subsection{Comprehensiveness of Bugs in \datasetns}

To understand whether bugs in \dataset are comprehensive and representative, we studied the bugs that were used to evaluate the existing resource leak detection techniques for Android apps~\cite{verifier}\cite{relda}\cite{relda2}\cite{greendroid_tse}\cite{elite}. We found that bug examples or patterns discussed in these studies are all observed in \datasetns. For example, Wu et al.~\cite{relda2} discussed resource leaks due to complex app component lifecycles. Vekris et al.~\cite{verifier} pointed out that the high level of asynchronous computing in Android apps (e.g., multi-threading) also often lead to resource leaks. As we can see from Section~\ref{ssec:leak_extent}, \dataset provides similar instances of such bug patterns and include many other types of bugs. Based on this observation, we believe that bugs in our \dataset are comprehensive and representative.

\subsection{Usefulness of \datasetns}
\dataset has many potential applications. We discuss several usage scenarios in the following.
\begin{itemize}[leftmargin=*,itemsep=4pt]
	\item{\textit{Evaluating bug detection tools.} \dataset contains a large number of resource leak bugs, covering diverse types of resources. These bugs can be used to evaluate existing resource leak detection tools for Android apps~\cite{studio}\cite{jdt}\cite{infer}\cite{relda}\cite{greendroid_tse}\cite{elite}\cite{verifier}\cite{relda2} to understand their strengths and limitations and thus guide the future development of similar techniques.}
	
	\item{\textit{Enabling resource leak patching research.} Automatically patching program defects can significantly improve the productivity of software developers. Since resource management policies are well-defined, it is possible to automatically fix potential resource leaks in programs to improve their performance and reliability. Towards this end, \dataset provides diverse resource leak bugs and human-written patches to facilitate the research in automated resource leak patching for Android apps.}
	
	\item{\textit{Supporting pattern-based bug detection.} Our empirical study revealed common patterns of faults made by Android developers. Such patterns can be leveraged to design static checkers (e.g., plug-ins to Lint~\cite{lint}) for real time detection of resource leaks in Android apps. We will demonstrate the feasibility of doing so in Section~\ref{sec:case_study}.}
\end{itemize}

\section{A Mini Case Study}\label{sec:case_study}
To investigate the usefulness of \datasetns, we conducted a mini case study to see whether our findings can help resource leak bug detection. Since we observed that Android developers often forget to release empty database cursors, we implemented a static checker on the Soot program analysis framework~\cite{soot} to detect the misuses of the \texttt{\small moveToFirst()} API (see Figure~\ref{fig:api_misuse}(a) for an example bug). The checker takes as input an Android app's Java bytecode (\texttt{\small .class} files) or \texttt{\small .apk} file as input and performs forward flow analysis for bug detection. Specifically, the checker analyzes each method and pinpoints the \texttt{\small if} statements whose condition involves checking the return value of the \texttt{\small moveToFirst()} method call on a cursor object \texttt{\small c} against \texttt{\small true} or \texttt{\small false} and identifies the branch where \texttt{\small moveToFirst()} evaluates to \texttt{\small false}. Then the checker further analyzes whether a statement that invokes the \texttt{\small close()} API on the cursor \texttt{\small c} is reachable from that branch. If yes, the check considers that the empty cursors are properly closed. Otherwise, the checker raises a warning.

\begin{table}
	\centering
	\caption{Detected Leaks of Database Cursors}
	\label{tab:results}
	\renewcommand{\arraystretch}{1.2}
	\renewcommand{\tabcolsep}{0pt}
	\begin{tabular}{|P{0.25\columnwidth}|P{0.25\columnwidth}|P{0.25\columnwidth}|P{0.25\columnwidth}|}
		\hline
		\textbf{App name} & \textbf{Version} & \textbf{\# found bugs} & \textbf{Bug report ID} \\ \hline
		ChatSecure & b598e57e8a & 1 & 755 \\ \hline
		IRCCloud & a96eda0860 & 2 & 147\textsuperscript{+} \\ \hline
		SureSpot & 76b6f931b0 & 3 & 142\textsuperscript{+} \\ \hline
		OwnCloud & b7577d8d86 & 1 & 1818\textsuperscript{+} \\ \hline
		SMSDroid & 20e9fb149b & 1 & 31\textsuperscript{+} \\ \hline 
		Osmand & 0970ad6496 & 1 & 3135\textsuperscript{+} \\ \hline 
		OSMTracker & d80dea16e4 & 2 & 74\textsuperscript{+} \\ \hline 
		OI File Manager & 03aa8903e2 & 1 & 82\textsuperscript{+} \\ \hline 
		\multirow{2}{*}{WordPress} & \multirow{2}{*}{4a90526c41} & 5 & 4591\textsuperscript{+}  \\ \cline{3-4}
		& & 1 & 4526*  \\ \hline
	\end{tabular}
\end{table}

\textit{Study results.} We applied the checker on the latest version of our 34 app subjects. Encouragingly, it successfully detected 18 database cursor leaks in nine apps. Table~\ref{tab:results} reports the names of these apps and their versions, in which our checker detected bugs. We then reported our findings to the developers of these apps. The last column of Table~\ref{tab:results} gives the ID of our bug reports. So far, 14 bugs reported in eight bug reports (marked with ``*'' or ``\textsuperscript{+}'')  have been confirmed by developers. Eight bugs have been quickly fixed by developers themselves or merging our pull requests (the bug reports are marked with ``\textsuperscript{+}''). Two bug reports are still pending (the issues were reported around two weeks before our paper submission and some apps are not actively maintained these days). Such results show the potential usefulness of \datasetns.

\section{Related Work}\label{sec:related_work}

\subsection{Resource Management}
System resources are finite. Developers are required to release resources used by their apps in a timely fashion when the resources are no longer needed. However, tasks for realizing this requirement are often error-prone. Empirical evidence shows that resource leaks are common in practice~\cite{weimer}. To prevent resource leaks, researchers proposed various language-level mechanisms and automated management techniques~\cite{closer}. Various tools were also developed to detect resource leaks~\cite{greendroid_tse}\cite{qvm}\cite{tracker}\cite{relda2}\cite{verifier}. For example, QVM~\cite{qvm} is a specialized runtime that tracks the execution of Java programs and checks for the violations of resource management policies. Tracker~\cite{tracker} is an industrial-strength tool for finding resource leaks in Java programs. These techniques are applicable to Android apps, which are typically Java programs, but they do not deal with the specialties in Android apps (e.g., implicit control flows). Therefore, in recent years, researchers also tailored resource leak detection techniques for Android apps. Examples are Relda~\cite{relda}, Relda2~\cite{relda2}, LeakDroid~\cite{testing}\cite{leakdroid}, and our earlier work GreenDroid~\cite{greendroid_tse}. Besides the efforts from research communities, there are also industrial tools for resource leak detection for Android apps, such as Facebook Infer~\cite{infer} and the built-in checkers in Eclipse~\cite{jdt} and Android Studio~\cite{studio}. Despite the existence of so many techniques, there does not exist a common set of real-world resource leak bugs in Android apps to facilitate the evaluation and comparison of these techniques. Our work makes the first attempt towards benchmarking resource leak bugs in Android apps.

\subsection{Memory Usage Analysis}
Programs written in the Java programming language enjoy the benefits of garbage collection, which frees the developers from the responsibility of memory management. Although developers do not need to care about explicitly recycling the created objects, memory leak may still happen when the programs maintain references to objects that prevent garbage collection or constantly create objects that have poor utility. To help diagnose such memory usage problems, many techniques have been developed. For example, researchers proposed to use object staleness~\cite{bell}\cite{profiling}, growing instances of types~\cite{cork}\cite{leakbot}, and cost-benefit analysis~\cite{cost-benefit} to identify suspicious and low-utility data structures that may cause memory leaks. Similar to resource leaks, besides tools originating from research communities, there are also industrial tools for memory usage analysis. For example, DDMS~\cite{ddms} in Android Studio and MAT~\cite{mat} in Eclipse are both powerful and fast tools to help Android developers analyze heap usage for finding memory leaks and reducing memory consumption. Many bugs in \dataset cause memory wastes and they can also be used to evaluate these techniques.

\subsection{Bug Benchmarking}
Bug benchmarks enable controlled experimentation and reproducible research. In early years, researchers constructed Siemens~\cite{siemens}, a widely-used benchmark of faulty programs. It provides a set of small to median sized C programs with manually seeded faults to facilitate the evaluation of dataflow and control flow based testing techniques. SIR~\cite{sir} is the first benchmark that contains real bugs in Java, C, C++, and C\# programs, but still the majority of the bugs were manually seeded or obtained by mutation and the program sizes are small. In recent years, researchers started to construct benchmarks of real bugs from large-scale software as many complex systems have been open-sourced~\cite{radbench}\cite{bugbench}\cite{defects4j}. One typical example is Defects4J~\cite{defects4j}. It provides 357 bugs from five large Java programs with exposing test cases. Compared to such benchmarks, our \dataset has its unique features. First, to the best of our knowledge, it is the first collection of real bugs in large-scale Android apps. Second, \dataset focuses on resource leaks and covers a wide range of different resource classes. Third, due to its focus, \dataset features resource leaks that occurred due to similar root causes and coding mistakes, which can support future research.

\section{Conclusion}\label{sec:conclusion}
This paper described \datasetns, a collection of \numbugs resource leak bugs in real-world Android apps. We constructed \dataset by analyzing 124,215 code revisions of 34 large-scale open-source app subjects. To understand the characteristics of these bugs, we conducted an empirical study on \dataset and discovered the root causes why resource leaks are common in Android apps and several patterns of faults made by developers. To show the usefulness of our study, we implemented a static checker to automatically find resource leak bugs that are similar to those in \datasetns. Experiments on representative Android apps showed that the checker can effectively detect resource leak bugs.

In future, we expect \dataset to further grow and contain more diverse bug instances. We also plan to evaluate the existing resource leak detection techniques using \dataset and quantitatively compare their strengths and weaknesses to see whether we can observe new challenges that need to be addressed for effective resource leak detection. With these efforts, we hope to shed light on future research and facilitate the development of effective automated resource leak detection and patching techniques.




\begin{thebibliography}{1}

\bibitem{anr}
Android ANR Errors.
\url{https://developer.android.com/training/articles/perf-anr.html}.
	
\bibitem{api}
Android API Guides.
\url{https://developer.android.com/guide/}.

\bibitem{process}
Android Processes and Threads.
\url{https://developer.android.com/guide/components/processes-and-threads.html}.

\bibitem{studio}
Android Studio's Built-in IntelliJ Code Inspection. \url{https://www.jetbrains.com/help/idea/2016.2/code-inspection.html}.

\bibitem{ankidroid}
AnkiDroid Code Repository.
\url{https://github.com/ankidroid/Anki-Android/}.

\bibitem{commons-io}
Apache Commons IO Library.
\url{http://commons.apache.org/proper/commons-io/}.

\bibitem{csipsimple}
CSipSimple Code Repository.
\url{https://github.com/r3gis3r/CSipSimple}

\bibitem{ddms}
DDMS: Dalvik Debug Monitor Server.
\url{https://developer.android.com/studio/profile/ddms.html}.
	
\bibitem{jdt}
Eclipse JDT Built-in Checkers. \url{http://help.eclipse.org/luna/topic/org.eclipse.jdt.doc.user/reference/preferences/java/compiler/ref-preferences-errors-warnings.htm}.

\bibitem{fbreader}
FBReaderJ Code Repository.
\url{https://github.com/geometer/FBReaderJ}.


\bibitem{fdroid}
F-Droid: A Catalogue of Open-Source Android Apps.
\url{https://f-droid.org/}.

\bibitem{fragments}
Fragments in Android Apps.
\url{https://developer.android.com/guide/components/fragments.html}.

\bibitem{github}
GitHub.
\url{https://github.com/}.


\bibitem{infer}
Infer: Bug Detection Tool for Android and iOS Apps. \url{http://fbinfer.com/}.

\bibitem{irccloud}
IRCCloud Code Repository.
\url{https://github.com/irccloud/android}.

\bibitem{mat}
MAT: Memory Analyzer in Eclipse.
\url{http://www.eclipse.org/mat/}.

\bibitem{leakbench}
\dataset: A Collection of Resource Leak Bugs in Android Apps.
\url{http://sccpu2.cse.ust.hk/droidleaks}.

\bibitem{lint}
Lint: Code Scanning Tool for Android apps.
\url{https://developer.android.com/studio/write/lint.html}.

\bibitem{sipdroid}
SipDroid Code Repository.
\url{https://github.com/i-p-tel/sipdroid}.

\bibitem{soot}
Soot: A Framework for Analyzing and Transforming Java and Android Apps.
\url{http://sable.github.io/soot/} 

\bibitem{qvm}
M. Arnold, M. Vechev, and E. Yahav, ``QVM: An Efficient Runtime for Detecting Defects in Deployed Systems,'' in ACM Transactions on Software Engineering Methodology (TOSEM), 2011, vol. 21, no. 1, pp. 2:1-2:35.

\bibitem{bell}
M. D. Bond and K. S. McKinley, ``Bell: Bit-Encoding Online Memory Leak Detection,'', in Proceedings of the 12th International Conference on Architectural Support for Programming Languages and Operating Systems (ASPLOS'06), 2006, pp. 61-72.

\bibitem{ibug}
V. Dallmeier and T. Zimmermann, ``Extraction of Bug Localization Benchmarks from History,'' in Proceedings of the 22nd International Conference on Automated Software Engineering (ASE'07), 2007, pp. 433-436.

\bibitem{closer}
I. Dillig, T. Dillig, E. Yahav, and S. Chandra, ``The CLOSER: Automating Resource Management in Java,'' in Proceedings of the 7th International Symposium on Memory Management (ISMM'08), 2008, pp. 1-10.

\bibitem{sir}
H. Do, S. Elbaum, and G. Rothermel, ``Supporting Controlled Experimentation with Testing Techniques: An Infrastructure and Its Potential Impact,'' in Empirical Software Engineering (EMSE), vol. 10, no. 4, pp. 405-435.

\bibitem{permission}
A. P. Felt, E. Chin, S. Hanna, D. Song, and D. Wagner, ``Android Permission Demystified,'' in Proceedings of the 18th ACM Conference on Computer and Communications Security (CCS'11), 2011, pp. 627-638.
  
\bibitem{relda}
C. Guo, J. Zhang, J. Yan, Z. Zhang, and Y. Zhang, ``Characterizing and Detecting Resource Leak,'' in Proceedings of the 28th IEEE/ACM International Conference on Automated Software Engineering (ASE'13), 2013, pp. 389-398.

\bibitem{profiling}
M. Hauswirth and T. M. Chilimbi, ``Low-Overhead Memory Leak Detection using Adaptive Statistical Profiling,'' in Proceedings of the 10th International Conference on Architectural Support for Programming Languages and Operating Systems (ASPLOS'04), 2004, pp. 156-164.

\bibitem{siemens}
M. Hutchins, H. Foster, T. Goradia, and T. Ostrand, ``Experiments of the Effectiveness of Dataflow- and Controlflow- Based Test Adequacy Criteria,'' in Proceedings of the 16th International Conference on Software Engineering (ICSE'94), 1994, pp. 191-200.

\bibitem{radbench}
N. Jalbert, C. Perira, G. Pokam, and K. Sen, ``RADBench: A Concurrency Bug Benchmark Suite,'' in Proceedings of the 3rd USENIX Conference on Hot Topic in Parallelism (HotPar'11), 2011, pp. 2:1-2:6.

\bibitem{cork}
M. Jump and K. S. McKinley, ``Cork: Dynamic Memory Leak Detection for Garbage-Collected Languages,'' in Proceedings of the 34th Annual ACM SIGPLAN-SIGACT Symposium on Principles of Programming Languages (POPL'07), 2007, pp. 31-38.

\bibitem{defects4j}
R. Just, D. Jalali, and M. D. Ernst, ``Defects4J: A Database of Existing Faults to Enable Controlled Testing Studies for Java Programs,'' in Proceedings of 23rd International Symposium on Software Testing and Analysis (ISSTA'14), 2014, pp. 437-440. 

\bibitem{async}
Y. Lin, C. Radoi, and D. Dig, ``Retrofitting Concurrency for Android Applications through Refactoring,'' in Proceedings of the 22nd ACM SIGSOFT International Symposium on Foundations of Software Engineering (FSE'14), 2014, pp. 341-352.

\bibitem{greendroid}
Y. Liu, C. Xu, and S.C. Cheung, ``Where Has My Battery Gone? Finding Sensor Related Energy Black Holes in Smartphone Applications,'' in Proceedings of the 11th IEEE International Conference on Pervasive Computing and Communications (PerCom'13), 2013, pp. 2-10.

\bibitem{greendroid_tse}
Y. Liu, C. Xu, S.C. Cheung, and J. Lu, ``GreenDroid: Automated Diagnosis of Energy Inefficiency for Smartphone Applications,'' in IEEE Transactions on Software Engineering, vol. 40, iss. 9, 2014, pp. 911-940.

\bibitem{elite}
Y. Liu, C. Xu, S.C. Cheung, and V. Terragni, ``Understanding and Detecting Wake Lock Misuses for Android Applications,'' in Proceedings of the 24th ACM SIGSOFT International Symposium on the Foundations of Software Engineering (FSE'16), 2016. 

\bibitem{stemming}
J. Lovins, ``Development of a Stemming Algorithm,'' in Mechanical Translation and Computational Linguistics, vol. 11, no. 1 and 2, 1968, pp. 22-31.

\bibitem{bugbench}
S. Lu, Z. Li, F. Qin, L. Tan, P. Zhou, and Y. Zhou, ``Bugbench: Benchmarks for Evaluating Bug Detection Tools,'' in Proceedings of the Workshop on the Evaluation of Software Defect Detection Tools (Bugs'05), 2005, pp. 1-5.

\bibitem{leakbot}
N. Mitchell and G. Sevitsky, ``Leakbot: An Automated and Lightweight Tool for Diagnosing Memory Leaks in Large Java Applications,'' in Proceedings of the 17th European Conference on Object-Oriented Programming (ECOOP'03), 2003, pp. 351-377.

\bibitem{tracker}
E. Torlak and S. Chandra, ``Effective Interprocedural Resource Leak Detection,'' in Proceedings of the 32nd ACM/IEEE International Conference on Software Engineering (ICSE'10), 2010, pp. 535-544.

\bibitem{verifier}
P. Vekris, R. Jhala, S. Lerner, and Y. Agarwal, ``Towards Verifying Android Apps for Absence of No-Sleep Energy Bugs,'' in Proceedings of the 2012 USENIX Conference on Power-Aware Computing and Systems (HotPower'12), 2012, article 3, pp. 1-5.

\bibitem{weimer}
W. Weimer and G. C. Necula, ``Finding and Preventing Run-Time Error Handling Mistakes,'' in Proceedings of the 19th annual ACM SIGPLAN Conference on Object-Oriented Programming, Systems, Languages, and Applications (OOPSLA'04), 2004, pp. 419-431. 

\bibitem{relda2}
T. Wu, J. Liu, Z. Xu, C. Guo, Y. Zhang, J. Yan, and J. Zhang, ``Light-weight, Inter-procedural and Callback-aware Resource Leak Detection for Android Apps,'' in IEEE Transactions on Software Engineering, 2016.

\bibitem{cost-benefit}
G. Xu, N. Mitchell, M. Arnold, A. Rountev, E. Schonberg, and G. Sevitsky, ``Finding Low-Utility Data Structures,'' in Proceedings of the 31st ACM SIGPLAN Conference on Programming Language Design and Implementation (PLDI'10), 2010, pp. 174-186.

\bibitem{testing}
D. Yan, S. Yang, and A. Rountev, ``Systematic Testing for Resource Leaks in Android Applications,'' in Proceedings of the 24th IEEE International Symposium on Software Reliability Engineering (ISSRE'13), 2013, pp. 411-420.

\bibitem{leakdroid}
H. Zhang, H. Wu, and A. Rountev, ``Automated Test Generation for Detection of Leaks in Android Applications'' in Proceedings of 11th IEEE/ACM International Workshop on Automation of Software Test (AST'16), 2016, pp. 64-70.

\end{thebibliography}
\end{document}